\documentclass[aip,apl,amsmath,amssymb,reprint]{revtex4-1}
\usepackage[T1]{fontenc}
\usepackage[utf8]{inputenc}
\usepackage{graphics,bm}
\usepackage{amssymb}
\usepackage{epsfig}
\usepackage{epsfig}
\usepackage{epsf}
\usepackage{graphicx}
\usepackage{amsmath}
\usepackage{mathrsfs}
\usepackage{amsfonts}
\usepackage{color}

\begin{document}
\title{Brownian motion of massive skyrmions forced by spin polarized currents }
\author{Roberto E. Troncoso}
\email{R.E.TroncosoCona@gmail.com}
\affiliation{ Centro para el Desarrollo de la Nanociencia y la Nanotecnolog\'ia, CEDENNA, Avda. Ecuador 3493, Santiago 9170124, Chile}
\author{Alvaro S. N\'u\~nez}
\affiliation{Departamento de F\'{i}sica, Facultad de Ciencias F\'{\i}sicas y Matem\'aticas, Universidad de Chile, Blanco Encalada 2008, Santiago, Chile}

\begin{abstract}
We report on the thermal effects on the motion of current-driven massive magnetic skyrmions. The reduced equation for the motion of skyrmion has the form of a stochastic generalized Thiele's equation. 
We propose an ansatz for the magnetization texture of a non-rigid single skyrmion that depends linearly with the velocity. By utilizing this ansatz it is found that the mass of skyrmion is closely related to intrinsic skyrmion parameters, such as Gilbert damping, skyrmion-charge and dissipative force. We have found an exact expression for the average drift velocity as well as the mean-square velocity of the skyrmion. The longitudinal and transverse mobility of skyrmions for small spin-velocity of electrons is also determined and found to be independent of the skyrmion mass.
\end{abstract}

\maketitle
%++++++++++++++++++++++++++++++++++++++++++++++++++++++++++++
{\em Introduction}.- Skyrmions have recently been the focus of intense research in spintronics. They are vortex-like spin structures that are topologically protected\cite{Skyrme,Bogdanov}. A series of works report their recent observation in chiral magnets\cite{Muhlbauer,Jonietz,Munzer,PMilde,Nagao,Seki}. There is a great interest in their dynamics due to the potential applications in spintronics that arise from the rather low current densities that are necessary to manipulate their location\cite{Fert}. Among other systems that have been reported hosting skyrmion textures they were observed in  bulk magnets MnSi\cite{Muhlbauer,Jonietz}, Fe$_{1-x}$Co$_x$Si\cite{Munzer,PMilde,Yu2}, Mn$_{1-x}$Fe$_x$Ge\cite{Shibata} and FeGe\cite{Yu1} by means of neutron scattering and Lorentz transmission electron microscopy. Regarding their dimensions, by the proper tuning of external magnetic fields, sizes of the order of a few tens of nanometers have been reported.  Spin transfer torques can be used to manipulate and even create isolated skyrmions in thin films as shown by numerical simulations\cite{Iwasaki,Iwasaki2,Lin2}. In thin films skyrmions have been observed at low temperatures, however energy estimates predict the stability of isolated skyrmions even at room temperatures\cite{Heinze}. Under that regime the motion of skyrmions is affected by fluctuating thermal torques that will render their trajectories into stochastic paths very much like the Brownian dynamics of a particle.    Proper understanding of such  brownian motion is a very important aspect of skyrmion dynamics.  Numerical simulations\cite{Kong1,Lin0} and experimental results\cite{nagaosa2}, suggest that the skyrmion position can be manipulated by exposure to a thermal gradient and that the skyrmions also display a thermal creep motion in a pinning potential\cite{Lin}. The thermally activated motion of pinned skyrmions has been studied in Ref. [\onlinecite{troncoso1}] where it has been reported that thermal torques can increase the mobility of skyrmions by several orders of magnitude. 
In this work we present a study of the random motion of magnetic skyrmions arising from thermal fluctuations. In our analysis we include an assessment of the deformation of the skyrmion that arises from its motion. This deformation induces an inertia-like term into the effective stochastic dynamics of the skyrmion. We present a theory that allows us to establish a relation between the fluctuating trajectory of the skyrmion and its effective mass. 

{\em Stochastic dynamics}.- We begin our analysis from the stochastic Landau-Lifschitz-Gilbert (LLG) equation\cite{Brown,Duine} that rules the dynamics of the magnetization direction ${\bf \Omega}$. Into this equation we need  to include the adiabatic, given by $-{\bf v}_s\cdot\nabla{\bf \Omega}$, and non-adiabatic, given by $\beta{\bf v}_s\cdot\nabla{\bf \Omega}$,  spin-transfer torques\cite{Berger, Slonczewski} where the strength of the non-adiabatic spin-transfer torque is characterized by the parameter  $\beta$. 
In those expressions ${\bf v}_s=-\left(pa^3/2eM\right){\bf j}$ stands for the spin-velocity of the conduction electrons, $p$ is the spin polarization of the electric current density ${\bf j}$, $e(>0)$ the elementary charge, $a$ the lattice constant, and $M$ the magnetization saturation. With those contributions the stochastic Landau-Lifshitz-Gilbert equation becomes:
\begin{align}\label{eq:LLG}
\left(\frac{\partial}{\partial t}+{\bf v}_s\cdot\nabla\right){\bf \Omega}={\bf \Omega}&\nonumber\times \left({\bf H}_{\text{eff}}+{\bf h}\right)\\
&+\alpha{\bf \Omega}\times\left(\frac{\partial}{\partial t}+\frac{\beta}{\alpha}{\bf v}_s\cdot\nabla\right){\bf \Omega},
\end{align}
where ${\bf H}_{\text{eff}}=\frac{1}{\hbar}\frac{\delta E}{\delta{\bf \Omega}}$ is the effective field, with $E$ representing the energy of the system, and $\alpha$ the Gilbert damping constant. An important aspect of this equation is the inclusion of the white Gaussian fluctuating magnetic field ${\bf h}$, describing the thermal agitation of the magnetization and obeying the fluctuation-dissipation theorem\cite{Brown}. The strength of the noise, $\sigma=2\alpha k_BTa^2/\hbar$, is proportional to the thermal energy $k_BT$, the Gilbert damping parameter $\alpha$, and the volume of the finite element grid $a^2$.
\\
Particle like solutions of the Landau-Lifshitz-Gilbert equation, that represent compact magnetic textures moving as coherent entities with a well defined velocity, have known since long ago. Among other examples we can found  the dynamics of domain walls \cite{hayashi,alvaro} and of Bloch points\cite{Gab,Hertel}. 
The account of the dynamics of skyrmion textures is best handled by the use of the collective coordinates approach. Under this framework the complex dynamics of the magnetization texture, ${\bf\Omega}({\bf r},t)$ is reduced to the evolution of a small number of degrees of freedom given by the skyrmion position and its velocity. In this way the magnetization field associated to a single-skyrmion moving along the trajectory ${\bf x}(t)$ is represented by a magnetization profile ${\bf \Omega}({\bf r},t)={\bf \Omega}({\bf r}-{\bf x}(t),{\bf v}(t))$. The explicit time-dependence of the magnetization, coming from the dependence on velocity ${\bf v}(t)$, includes the effects of deformations of the skyrmion\cite{Moutafis,Makhfudz,Moon}. The calculation for the static skyrmion profile, ${\bf \Omega}_0({\bf r})$, has been addressed elsewhere\cite{Knoester}, by means of a minimization of the magnetic energy. In this energy the contributions from the exchange, perpendicular anisotropy, and Dzyaloshinskii-Moriya energies must be included. Replacement of the collective coordinates ansatz and integration over space reduce the LLG equation to an equation of motion for the collective variables. This equation has the form of a stochastic massive Thiele's equation
\begin{align}\label{eq:stochasticThiele}
{\mathcal {M}}\dot{\bf v}(t)-g\hat{z}\times {\bf v}(t)+\alpha{\cal D}{\bf v}(t)={\bf F}+{\boldsymbol{\eta}}(t).
\end{align}

Neglecting the contribution of the noise term (${\boldsymbol{\eta}}(t)$) Eq. (\ref{eq:stochasticThiele}) turns into  the  generalized Thiele's equation \cite{Thiele}. We highlight the inertial terms, quantified by the effective mass, that correspond to a matrix ${\mathcal M}_{ij}={M}_{ij}+{\bar M}_{ij}$, comprised by the elements $M_{ij}=\int d{\bf r}\;{\bf \Omega}\cdot\left(\frac{\partial\bf \Omega}{\partial x^{i}}\times\frac{\partial\bf \Omega}{\partial v^{j}}\right)$, arising from the conservative dynamics, and ${\bar M}_{ij}=\alpha\int d{\bf r}\;\left(\frac{\partial\bf \Omega}{\partial x^{i}}\cdot\frac{\partial\bf \Omega}{\partial v^{j}}\right)$ arising from the dissipative contribution to the Landau-Lifshitz-Gilbert equation. The second term in Eq. (\ref{eq:stochasticThiele}) describes the Magnus force \cite{Jonietz} exerted by the magnetic texture on the moving skyrmion.  In the case of an isolated skyrmion $g=4\pi W$ where $W=-1$ stands for the winding number, or skyrmion charge. The third contribution represents the dissipative force which is defined through the relation ${\cal D}_{ij}=\int d{\bf r}\frac{\partial{\bf \Omega}}{\partial x^i}\cdot\frac{\partial{\bf \Omega}}{\partial x^j}$, that becomes  ${\cal D}_{ij}={\cal D}\delta_{ij}$ in the case of the highly symmetrical case of an isolated skyrmion.

\begin{figure}[ht]
\begin{minipage}[b]{0.44\linewidth}
\centering
\includegraphics[width=1.56in,height=1.0in]{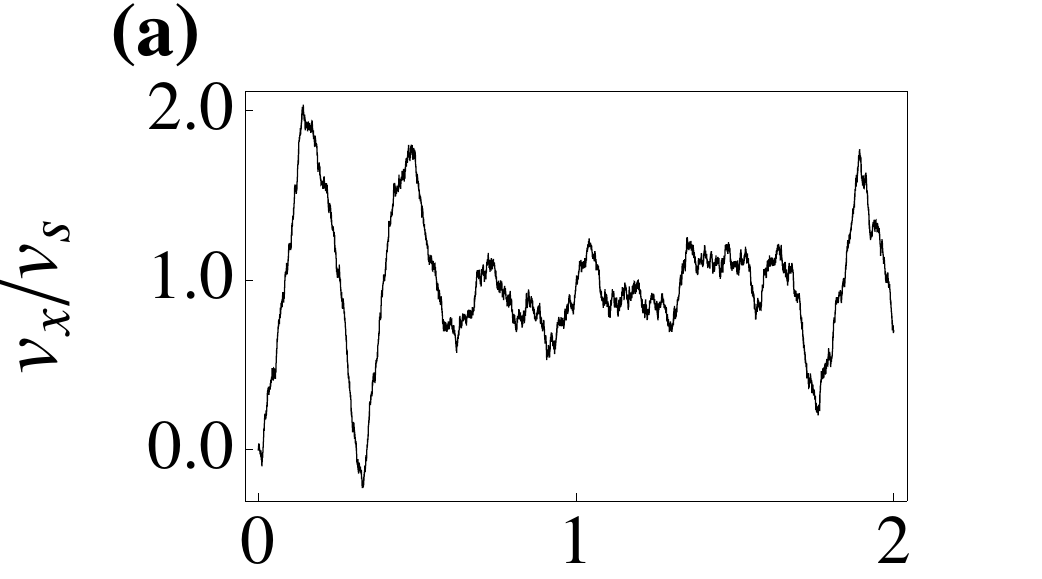}
\vspace{0.ex}
\includegraphics[width=1.55in,height=1.09in]{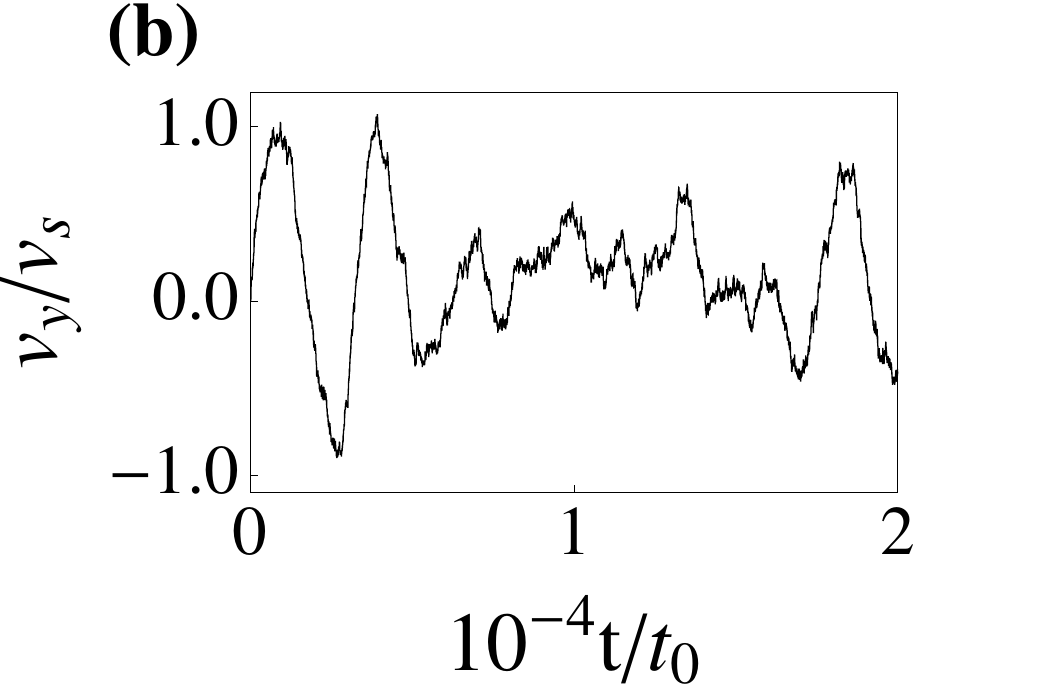}
\end{minipage}
\hspace{0.01cm}
\begin{minipage}[b]{0.535\linewidth}
\includegraphics[width=2.in,height=1.5in]{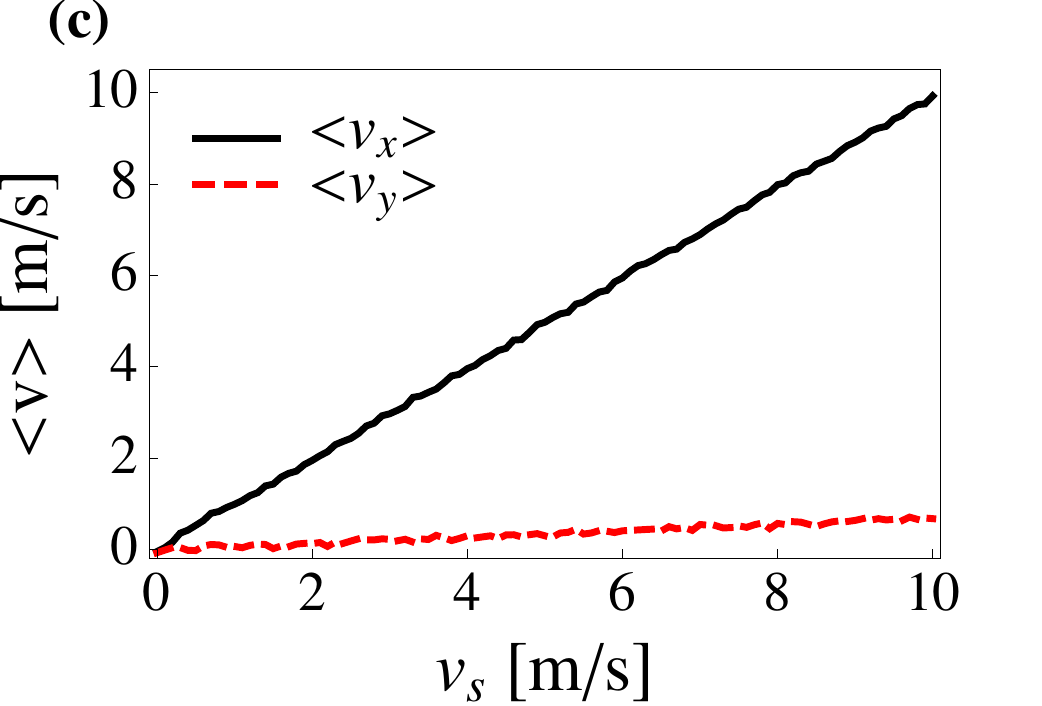}
\vspace{2.3ex}
\end{minipage}
\caption{Fluctuating skyrmion velocities in the longitudinal and transverse directions, (a) and (b) respectively. The Brownian dynamics Eq. (\ref{eq:stochasticThiele}) is solved for a spin velocity of electrons $v_s=1$ m/s along $x$ direction and for a characteristic time scale $t_0={\cal M}_{xx}\approx 6$ ns (from Ref. [\onlinecite{Buttner}]). In (c) the average velocities is presented as function of spin-velocity ${\bf v}_s=v_s\hat{x}$, both the longitudinal (black line) and transverse (dashed line) components. The results are shown for a Gilbert damping $\alpha=0.1$, $\beta=0.5\alpha$, and at temperature $100$ K.}
\label{fig:skyrmionvelocities}
\end{figure}

The dynamics of the skyrmion in  Eq. (\ref{eq:stochasticThiele}) is forced by a deterministic term ${\bf F}=-g{\hat z}\times{\bf v}_s+\beta {\cal D}{\bf v}_s-{\boldsymbol \nabla} V$, that contains a contribution from the flowing electrons and a force arising from a potential $V[{\bf x}]$ that reflects the inhomogeneities in the skyrmion`s path, e.g., magnetic impurities, local anisotropies or geometric defects.  We conclude with the last term of right-hand side of Eq. (\ref{eq:stochasticThiele}), that describes the fluctuating force on the skyrmion. The strength of the Gaussian white noise turns out to be $\sigma{\cal D}$, therefore the effective diffusion constant of the skyrmion depends not only on the Gilbert damping and the temperature but also on the dissipative parameter ${\cal D}$.
By solving the stochastic Thiele equation Eq. (\ref{eq:stochasticThiele}), for the homogenous case ($V[{\bf x}]=0$), we determine the time evolution of both longitudinal and transverse components of the velocity of the fluctuating skyrmion (as shown in Fig. \ref{fig:skyrmionvelocities} (a) and (b) respectively) at temperature $T=100$ K. Typical skyrmion speeds of $\sim 0-1$ m/s are reached for spin-velocities on the order of $1$ m/s. The massive skyrmion dynamics was calculated for a Gilbert-damping parameter $\alpha= 0.1$, the $\beta$ parameter $\beta=0.5\alpha$, the dissipative force ${\cal D} = 5.577\pi$ (from Ref. [\onlinecite{Iwasaki}]), and where the values used for the mass are taken from Ref. [\onlinecite{Buttner}]. Moreover, it is numerically solved the mean drift velocity, i.e., the steady-state current-induced skyrmion motion, which is displayed in Fig. \ref{fig:skyrmionvelocities} (c) as a function of the spin-velocity ${\bf v}_s=v_s\hat{x}$ of electrons.
In addition, we are interested in the probability distribution $P[{\bf x},{\bf v};t]$ associated to the skyrmion dynamics, which is defined as the probability density that a skyrmion at time $t$, is in the position ${\bf x}$ with a velocity ${\bf v}$. The equation of motion satisfied by such distribution is known as Fokker-Planck equation and its derivation, as well as its solution in simple cases, constitutes a standard issue in the analysis of stochastic processes\cite{Risken}.  

%%%%
{\em Origin of the skyrmion mass}.- The Brownian skyrmion motion described by Eq. (\ref{eq:stochasticThiele}) contains as a main ingredient the inertia term, regarding to Ref. [\onlinecite{troncoso1}], which is quantified by the effective mass matrix $\mathcal{M}$. Generally speaking, it is linked to the explicit time-dependence of the magnetization direction. The mass of skyrmions can be determined perturbatively in linear response theory as follows. In the skyrmion center of mass reference frame we see that the magnetization in Eq. (\ref{eq:LLG}) is affected by an additional magnetic field  $\delta{\boldsymbol{H}}({\bf r})={\bf\Omega}({\bf r})\times({\bf v}\cdot\nabla){\bf\Omega}({\bf r})$. We note that the strength of the effective field is confined within the perimeter of the skyrmion.  However, the effective torque takes a maximum value in the center of skyrmion and thus, it suffers a distortion of its shape due to the current-induced motion. This motivates us to propose an  ansatz for the magnetization texture of a non-rigid single skyrmion and its dependence on the velocity as
\begin{align}\label{eq:skyransatz}
{\bf \Omega}({\bf r},{\bf v})={\bf\Omega}_0({\bf r})+ \lambda\;\xi\; {\bf\Omega}_0({\bf r})\times({\bf v}\cdot\nabla){\bf\Omega}_0({\bf r}),
\end{align}
with ${\bf\Omega}_0$ corresponds to the static and rigid skyrmion texture. The deformation of skyrmion size is parameterized by the second term, where $\lambda$ is the dimensionless  parameter that determines the strength of the velocity induced deformations. In this expression $\xi=\hbar{l^2_{sk}}/{Ja^2} $ with $J$ the exchange constant, $a$ the lattice constant and $l_{sk}$ the characteristic skyrmion size. 
\begin{figure}[ht]
\begin{center}
\includegraphics[width=2.50in]{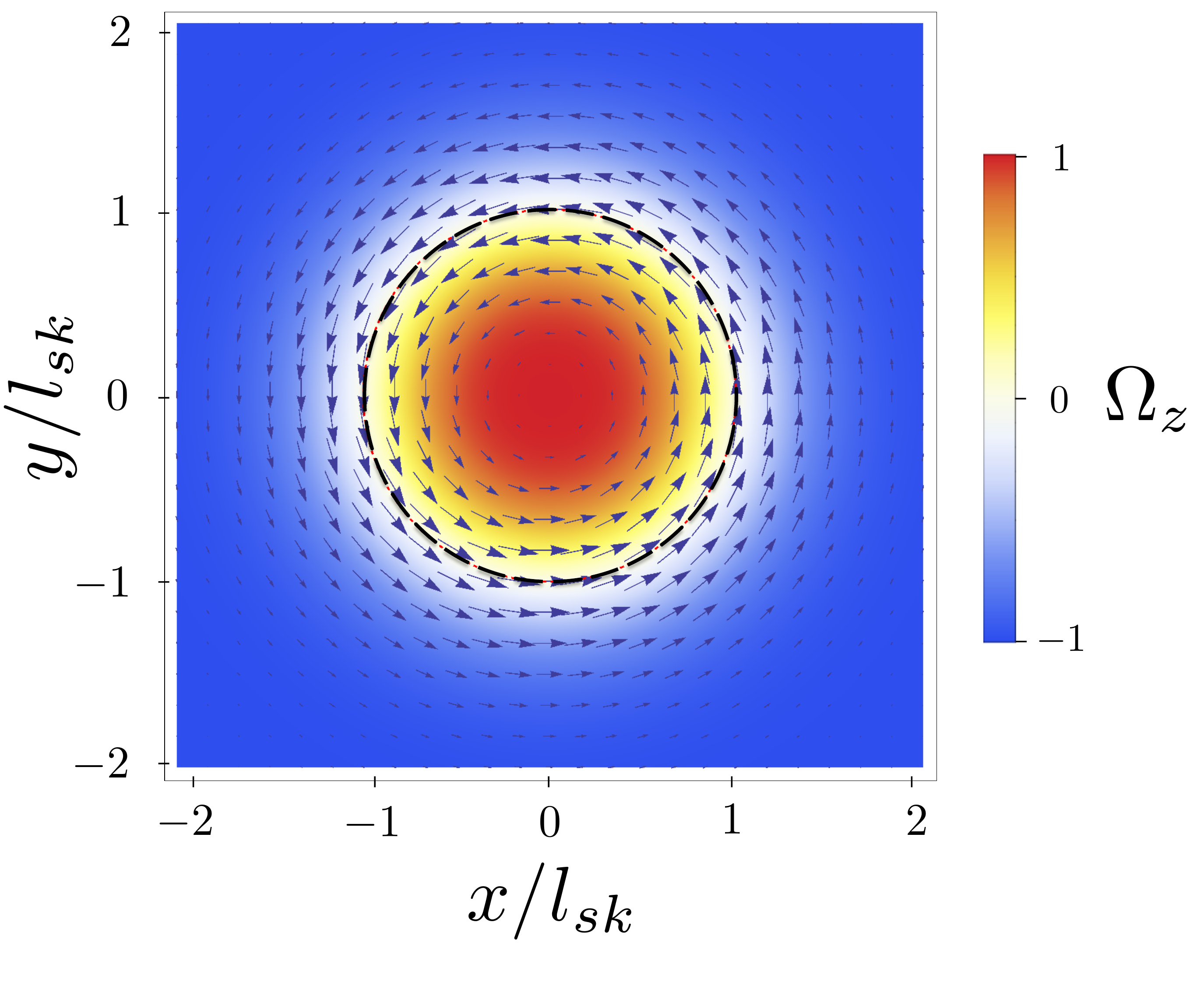}\\
\includegraphics[width=2.5in]{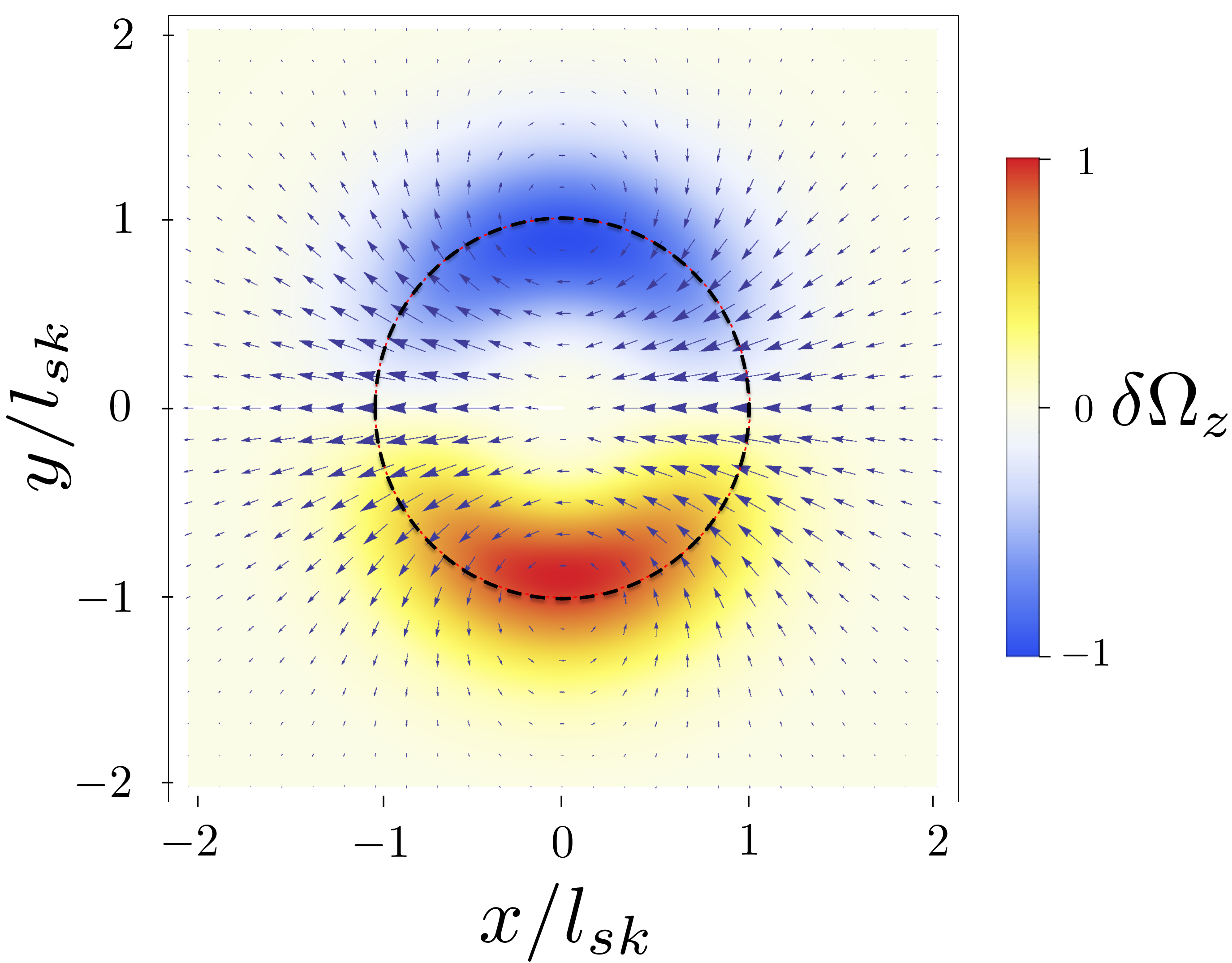}
\caption{Top: Pictorial representation of the skyrmion magnetic texture. The color encodes the out of plane component of the magnetization. It changes from being fully aling with the $+\hat{z}$ direction in the center to a complete alignment with the opposite direction in the outer rim. The arrows represent the behavior of the in plane component of the magnetization. For the case displayed those components swirl like a vortex. Bottom: Schematic plot for the skyrmion distortion generated by the motion of the skyrmion. In color we have represented the out of plane component of the deformation $\delta\Omega_z$. This contribution is concentrated in the direction transverse to the motion nearby the perimeter of the skyrmion (indicated by a dashed line). The arrows correspond to the in plane $(\delta\Omega_x,\delta\Omega_y)$.}
\label{fig:skyrmiondistortion}
\end{center}
\end{figure}
It is worth noting that this contribution is linear in velocity and conserves the norm of the magnetization to leading order in $\lambda$. In Fig. \ref{fig:skyrmiondistortion} we present schematically the distortion of skyrmions exerted by the effective field $\delta{\boldsymbol{H}}$. However, without loss of generality, we assume a motion of the skyrmion along $x$ direction. The deformation in the skyrmion texture consists of an in plane distortion, that resembles a dipolar field, and an out of plane contribution that is antisymmetric. The nature of the mass of skyrmions can be traced back by using the ansatz given by Eq. (\ref{eq:skyransatz}). By replacing it on the expressions for ${\cal M}$ up to linear order in $\lambda$, we find that the mass is related both to the dissipative matrix and gyrotropic tensor by ${\cal M}_{xx}={\cal M}_{yy}=\lambda\xi\mathcal{D}$ and ${\cal M}_{xy}=-{\cal M}_{yx}=\lambda\xi \alpha g$. We see how the dissipation mechanisms encoded by the Gilbert damping $\alpha$ generate an anti symmetrical contribution to the mass. By comparing our results for ${\cal M}_{xx}$ with earlier theoretical\cite{Moutafis,Makhfudz} and experimental\cite{Buttner} estimates of the skyrmion mass we obtain $\lambda\sim 0.01$. In this case  we obtain,  using a typical skyrmion velocity of 1 m/s in Eq. (\ref{eq:skyransatz}), a characteristic strength of the deformation of the skyrmion in the range of $|\delta{\bf \Omega}|\sim 10^{-3}$.

%%%%%% 
{\em Skyrmion mobility}.- The skyrmion dynamics is induced by an electric current density via spin-transfer torque mechanism. As mentioned previously, the flow of electrons exerts a force ${\bf F}$ that drives the skyrmion and its response is described by the stochastic generalized Thiele's equation\cite{troncoso1}. 
Next, we discuss the role of mass in the dynamics of skyrmions induced by currents at finite temperature. The probability distribution for the velocity of the skyrmion can be readily found for $V({\bf x})=0$ by solving the associated Fokker-Planck equation, $P[{\bf v}]=\mathcal{N} \exp\left(-\frac{1}{2}\gamma({\bf v}-{\bf \bar{v}})^2\right)$, where $\gamma=\lambda\xi \alpha (g^2+\mathcal{D}^2)/\sigma{\cal D}$ and $\bf \bar{v}=\mu_\parallel \bf v_s+\mu_\perp \hat{z}\times\bf v_s$. In this equation $\mu_\parallel= (g^2+\alpha\beta\mathcal{D}^2)/(g^2+\alpha^2\mathcal{D}^2)$ and $\mu_\perp=-(\alpha-\beta)g\mathcal{D}/(g^2+\alpha^2\mathcal{D}^2)$ are the longitudinal and transverse skyrmion mobilities respectively. We are interested in the average drift velocity, which is determined directly from the probability distribution as $\langle v_i\rangle=\bar{v}_i$. Another relevant quantity as the mean-squared velocity is evaluated directly, and is found to obey $\langle v_i v_j\rangle-\langle v_i\rangle\langle v_j\rangle =\sigma{\cal D} \left(\lambda\xi\alpha \left(g^2+\mathcal{D}^2\right)\right)^{-1}\delta_{ij}$ that relates the fluctuations on the velocity in terms of temperature and the skyrmion mass.  The mean drift velocity scales linearly with the spin-velocity of electrons and, unlike the current-induced domain wall dynamics\cite{alvaro}, skyrmions do not exhibit an intrinsic pinning\cite{Iwasaki,Iwasaki2}. Note that these values for the spin-velocities correspond to electric current densities on the order of $10^{10}$ A/m$^2$. As we analytically show, the average velocity of the free skyrmion is independent of the mass. However, it is related only on intrinsic parameters, such as Gilbert damping, skyrmion-charge, and dissipative force. Following our results, a detailed  characterization of the strength of velocity fluctuations can be used to determine the values of the skyrmion mass.

%%%%
{\em Conclusions}.- We have investigated the mechanisms by which thermal fluctuations influence the current-driven massive skyrmion dynamics. Based on the stochastic Landau-Lifschitz-Gilbert equation we derived the Langevin equation for the skyrmion. The equation has the form of a stochastic generalized Thiele's equation that describes the massive dynamics of a single-skyrmion at finite temperature. We proposed an ansatz for the magnetization texture of a non-rigid single skyrmion that depends linearly with the velocity. This ansatz has been derived based upon an effective field that distorts the skyrmion texture. In particular, it implies that the deformation of the skyrmion shape consists of an in plane distortion and an out of plane contribution that is antisymmetric. Furthermore, by utilizing this ansatz it is found that the mass of skyrmion is related with intrinsic parameters, such as Gilbert damping, skyrmion-charge, and dissipative force. This simple results provides a path for a theoretical calculation of the skyrmion mass.
We have found an exact expression for the average drift velocity as well as the mean-square velocity of the skyrmion. The longitudinal and transverse mobilities of skyrmions for small spin-velocity of electrons were also determined. We showed that the average velocity of skyrmions, unlike the mean-square velocity, is independent of the mass and it varies linearly with the spin-velocity. In future work, we plan to use the formalism developed in this work to the study of the transport of massive skyrmions in disordered media.
 
{\em Acknowledgements}.- The authors acknowledge funding from Proyecto Basal FB0807-CEDENNA, Anillo de Ciencia y Tecnonolog\'ia ACT 1117, and by N\'ucleo Cient\'ifico Milenio P06022-F.

\bibliographystyle{elsarticle-num}

\end{document}